# An Easy-to-Implement Hierarchical Standardization for Variable Selection Under Strong Heredity Constraint

Kedong Chen[1] · William Li[2] · Sijian Wang[3]

## Abstract
For many practical problems, the regression models follow the strong heredity property (also known as the marginality), which means they include parent main effects when a second-order effect is present. Existing methods rely mostly on special penalty functions or algorithms to enforce the strong heredity in variable selection. We propose a novel hierarchical standardization procedure to maintain strong heredity in variable selection. Our method is effortless to implement and is applicable to any variable selection method for any type of regression. The performance of the hierarchical standardization is comparable to that of the regular standardization. We also provide robustness checks and real data analysis to illustrate the merits of our method.

✉ William Li
wlli@saif.sjtu.edu.cn

Kedong Chen
kchen@odu.edu

Sijian Wang
sijian.wang@stat.rutgers.edu

[1] Department of Information Technology and Decision Sciences, Old Dominion University, Norfolk, USA

[2] Shanghai Advanced Institute of Finance, Shanghai Jiao Tong University, Shanghai, China

[3] Department of Statistics and Biostatistics, Rutgers University, Piscataway, USA

## 1 Introduction

Variable selection is important when data contain a large number of predictors. We often want to determine a smaller subset that exhibits the strongest effects. Numerous methods of variable selection have been proposed, such as the best subset selection, stepwise regression, and penalized regression including the lasso [26], the smoothly clipped absolute deviations (SCAD) [8], the minimax concave penalty (MCP) [31], and others.

In this paper, we consider the variable selection problem in the following two linear regression models:

$$Y = \eta + \sum_{j=1}^{p} \alpha_j X_j + \varepsilon, \tag{1}$$

$$Y = \eta + \sum_{j=1}^{p} \alpha_j X_j + \sum_{j=1}^{p} \beta_j X_j^2 + \sum_{1 \le j < k \le p} \gamma_{jk} X_j X_k + \varepsilon, \tag{2}$$

where $X_1, \ldots, X_p$ are the $p$ predictors, $Y$ is the response, $\eta$ is the intercept, $\alpha$'s, $\beta$'s, and $\gamma$'s are the coefficients, and $\varepsilon$ is the error term.

We distinguish between Model (1) and Model (2) because the second-order effects are necessary for many practical problems. For example, in statistical genomics, after identifying several disease-associated polymorphisms via the whole genome association analysis, researchers' aim is to detect effects that, due to the interaction among genetic (or environmental) factors, may not be identified using models containing only the first-order effects (i.e., main effects) [27]. In the oil extraction process, as another example, particle size, heating temperature, heating time, applied pressure and duration of pressing can jointly affect the yield and quality of mechanically expressed groundnut oil [1]. Two- or more-factor interactions are significant because the application environment of the product or process incorporates many factors simultaneously rather than separately at different times. Besides interaction terms, quadratic terms are often added to a linear regression model to study the curvature of the model.

When the second-order effects are present in a regression model, the principle of *effect heredity* is usually assumed [3, 10]. In this paper, we focus on the *strong effect heredity* (hereafter the "strong heredity") for which if a higher-order term $X_i X_j$ is included in the model, all corresponding parent main effects $X_i$ and $X_j$ are also included in the model.

Heredity, sparsity and hierarchy are the three main principles of experimental design [10]. Li et al. [18] provided empirical evidence to support these principles in practice. They analyzed a large collection of 113 experimental designs reported in the literature. With respect to heredity, they found that a two-factor interaction (2fi) had a probability of 33% of being significant if both of its parent main effects were also significant. In comparison, this probability declined to 4.5% if one of the two main effects was significant and to 0.48% if neither of them was significant.

The importance of strong heredity is also evident in a more general model selection framework. In a recent paper, Hao and Zhang [12] addressed some critical issues in the variable selection for linear models with interaction. They explained why the strong heredity condition is not that restrictive in practice, and they discussed the importance of following the *marginality principle*, which was first proposed by Nelder [20]. The marginality principle requires any interaction to be considered for selection only after its parents have entered the model, which is closely related to the strong heredity condition here.

Most of the classical variable selection methods do not guarantee heredity for the selected model. To address this limitation, for polynomial regressions, Peixoto [21, 22] considered hierarchical variable selection problems. Peixoto [22] illustrated the importance of using a well-formulated polynomial regression model (i.e., model that satisfies the strong heredity) in terms of having an invariant estimation space, but did not consider how to select the well-formulated model too much. The paper suggested that when considering the variable selection, only well-formulated models should be compared. This strategy works for the best-subset typed variable selection method, i.e., we can list all well-formulated models and calculate the $R^2$, $C_p$ or other statistics to pick up the best model. However, for the lasso-like penalized regression, this strategy does not work, because we cannot enforce the lasso to consider well-formulated models only.

From an alternative perspective, some researchers have proposed specially designed penalty functions and performed the *structured variable selection* to enforce the heredity. For example, Hamada and Wu [10] developed a modified stepwise procedure that can enforce heredity. Joseph and Delaney [14] proposed functionally induced priors (a Bayesian approach) that incorporate principles of effect hierarchy and effect heredity. Zhao et al. [32] introduced the composite absolute penalties (CAP) family to allow predictors to express grouping and hierarchical relationships. Based on [2], Yuan et al. [30] proposed novel non-negative garrote methods that can naturally incorporate hierarchical or structural relationships defined through effect heredity. Furthermore, Choi et al. [4] extended the lasso method [26] to the strong heredity interaction model (SHIM) through a novel reparameterization of the coefficients of the 2fi's. In the field of design of experiments (DOE), Errore et al. [7] illustrated the ability of definitive screening designs to correctly identify first- and second-order model terms through employing strong-heredity enforced methods like SHIM.

As noted in Hao and Zhang [12], two main types of interaction selection procedures are one-stage methods and two-stage methods. The former one involves selecting main and interaction effects simultaneously, subject to a hierarchical constraint. In comparison, two-stage methods, such as the two-stage least angle regression (LARS) [6], involve selecting main effects first and subsequently considering only interactions of those main effects. Yuan et al. [29] extended the LARS algorithm to address the effect heredity. Later, Hao and Zhang [12] justified the theoretical validity of the two-stage methods, claiming that such methods intentionally leave out interaction effects at stage one, leaving the model mis-specified. Their "iFORM" method [11] focuses more on the interaction selection from a high-dimensional perspective.

Compared to many of the methods proposed in the literature, we tackled the problem of enforcing heredity in the variable selection from a different angle. Instead of designing new penalty functions, we proposed a *hierarchical standardization* procedure without modifying the variable selection techniques, which is different from the existing standardization procedures. Standardization is a routine procedure in variable selection that makes all variables in the model on the same scale. Existing methods standardize the main effects, 2fi's, and quadratic terms in the same way. In comparison, our hierarchical standardization treats them differently. Following the hierarchical standardization, any variable selection method can be applied. When the estimated coefficients are transformed back to their original scales, the main effects will have nonzero estimated coefficients if all related 2fi's and/or quadratic terms have nonzero estimated coefficients. Any estimated coefficient that is nonzero is considered to be identified as important. In this sense, the heredity property is automatically guaranteed.

Compared to the heredity-guaranteed variable selection methods using special penalty functions, our proposed approach is much more flexible and adaptive. Users can apply the hierarchical standardization effortlessly with little or no change to the variable selection method. Moreover, some structured variable selection methods (e.g., SHIM) need to solve a non-convex optimization problem, which may cost much numerical effort and cause instability of the final solution in practice. By contrast, following our hierarchical standardization, if a convex variable selection method is used, the resulting heredity-guaranteed method is still convex.

The rest of this paper is organized as follows. We introduce the hierarchical standardization and the heredity-guaranteed variable selection framework in Sect. 2. In Sect. 3, we present simulation studies to show the merits of the proposed approach. We also conduct robustness checks on the framework through numerical studies. In Sect. 4, real data analysis illustrates the proposed strategy. We conclude the paper with a discussion in Sect. 5.

## 2 Hierarchical Standardization

Standardization is an important step for any penalized-estimation-based variable selection method, because the penalty function is not scale free. Typically, three steps are involved: (1) standardization of all variables in the model, (2) application of a variable selection method to the standardized data to obtain the estimated standardized coefficients and (3) transformation of the standardized coefficients to their original scales.

A commonly used standardization procedure in the literature, which we denote as *regular standardization* [13], makes each variable in the model have zero-mean and unit-standard deviation. For Model (2), the following equations show the correspondence between the unstandardized coefficients and the standardized coefficients with respect to the regular standardization. Transformation with respect to regular standardization:

$$\tilde{X}_j = \frac{X_j - \bar{X}_j}{s_j}, \quad \widetilde{X_j^2} = \frac{X_j^2 - \bar{X}_{jj}}{s_{jj}}, \quad \widetilde{X_j X_k} = \frac{X_j X_k - \bar{X}_{jk}}{s_{jk}} \quad (3)$$

Estimated model after regular standardization:

$$\hat{Y} = \tilde{\eta}^R + \sum_{j=1}^{p} \tilde{\alpha}_j^R \tilde{X}_j + \sum_{j=1}^{p} \tilde{\beta}_j^R \widetilde{X_j^2} + \sum_{1 \leq j < k \leq p} \tilde{\gamma}_{jk}^R \widetilde{X_j X_k} \quad (4)$$

Estimated original coefficients under regular standardization:

$$\hat{\alpha}_j^R = \frac{\tilde{\alpha}_j^R}{s_j}, \quad \hat{\beta}_j^R = \frac{\tilde{\beta}_j^R}{s_{jj}}, \quad \hat{\gamma}_{jk}^R = \frac{\tilde{\gamma}_{jk}^R}{s_{jk}}, \quad (5)$$

where $\bar{X}_j, \bar{X}_{jj}, \bar{X}_{jk}$ and $s_j, s_{jj}, s_{jk}$ are sample means and sample standard deviations of $X_j$, $X_j^2$, and $X_j X_k$, respectively; $\tilde{\alpha}^R$'s, $\tilde{\beta}^R$'s, and $\tilde{\gamma}^R$'s are the estimated standardized coefficients under regular standardization; $\hat{\alpha}^R$'s, $\hat{\beta}^R$'s, and $\hat{\gamma}^R$'s are the estimated original coefficients.

Except for some areas, such as the DOE area, the regular standardization has been the most commonly used method. The rationale, as suggested in the literature, is to keep variables on the same scale (see, for example, a popular book of James et al. [13, p. 217]. However, this procedure ignores the fact that quadratic effects and interactions are generated from main effects. Consequently, the selection of main effects, say $X_j$ and $X_k$, is not related to whether their 2fi $X_{jk}$ is selected. Hence, there is no guarantee that the final model satisfies the heredity constraint.

Different from the regular standardization, the proposed hierarchical standardization utilizes the hierarchical structure among model terms. As noted by an anonymous reviewer, such standardization has already been used in some commercial software such as JMP. After the main effects are standardized in the usual way (i.e., zero-mean and unit-standard deviation), the second-order terms are *generated* from the corresponding main effects, as illustrated in the following equations. Transformation with respect to hierarchical standardization:

$$\tilde{X}_j = \frac{X_j - \bar{X}_j}{s_j} \quad (6)$$

Estimated model after hierarchical standardization:

$$\hat{Y} = \tilde{\eta}^H + \sum_{j=1}^{p} \tilde{\alpha}_j^H \tilde{X}_j + \sum_{j=1}^{p} \tilde{\beta}_j^H \tilde{X}_j^2 + \sum_{1 \leq j < k \leq p} \tilde{\gamma}_{jk}^H \tilde{X}_j \tilde{X}_k \quad (7)$$

Estimated original coefficients under hierarchical standardization:

$$\hat{\alpha}_j^H = \frac{\tilde{\alpha}_j^H}{s_j} - 2\frac{\bar{X}_j \tilde{\beta}_j^H}{s_j^2} - \sum_{\substack{k \neq j \\ 1 \leq k \leq p}} \frac{\bar{X}_k \tilde{\gamma}_{jk}^H}{s_j s_k}, \quad \hat{\beta}_j^H = \frac{\tilde{\beta}_j^H}{s_j^2}, \quad \hat{\gamma}_{jk}^H = \frac{\tilde{\gamma}_{jk}^H}{s_j s_k}, \quad (8)$$

where the superscript $H$ stands for hierarchical standardization and the symbols carry similar meaning as in Eqs. (3)–(5). We provide an algorithmic procedure in "Appendix 2" to implement the hierarchical standardization.

Applying the hierarchical standardization guarantees strong heredity. In other words, after the estimated coefficients are transformed back to their original scales, the main effects have nonzero estimated coefficients, as long as all related 2fi's and/or quadratic terms have nonzero estimated coefficients. "Appendix 1" provides the proof for Eq. (8) and shows how the strong heredity is guaranteed. In general, if higher-order terms are selected and

$$\frac{\tilde{\alpha}_j^H}{s_j} - 2\frac{\bar{X}_j \tilde{\beta}_j^H}{s_j^2} - \sum_{\substack{k \neq j \\ 1 \leq k \leq p}} \frac{\bar{X}_k \tilde{\gamma}_{jk}^H}{s_j s_k} \neq 0 \quad (9)$$

then all the parent main effects will be guaranteed to be selected (see the proof in "Appendix 1"). For almost all observational studies, the entire $\bar{X}$ does not equal to $\mathbf{0}$. Hence, $\tilde{\beta}_j^H \neq 0$ and/or $\tilde{\gamma}_{jk}^H \neq 0$ represent $\hat{\alpha}_j^H \neq 0$. For designed experiments where $\bar{X}_i$ ($i = 1, \ldots, p$) is often set to zero, we can introduce a small shift $\delta \neq 0$, such that $\bar{X}_i = \delta$, to ensure the satisfaction of Eq. (9).

Hence, the hierarchical standardization guarantees that the estimated original coefficient of $X_j$ is not equal to 0 when the corresponding second-order effects are not equal to 0. In other words, due to the further back-transformation of the selected model, the updated selected model [Eq. (8)] maintains the strong heredity while the selected model itself, i.e., Eq. (7), may not maintain the strong heredity. We first select the model that best fits the criteria and then adjust the main effect coefficients to ensure strong heredity without significant deterioration in performance or accuracy (demonstrated in Sect. 3). Existing variable selection function or computing package (e.g., in R) can have $\widetilde{X_j^2}$ and $\widetilde{X_j X_k}$ as input variables generated from the hierarchical standardization [i.e., Eq. (6)].

Noticeably, there are many different ways to standardize the main effects for the first step of hierarchical standardization. Equation (6) shows the mean-standard deviation standardization, a most commonly used way that results in zero-mean and unit-standard deviation. Other widely used standardization includes the median-IQR (inter-quartile range) standardization, which might sometimes be preferable when the variables are skewed (see Sect. 3.4.2 for a numeric example). Nonetheless, regardless of the main-effect standardization used, the proposed hierarchical standardization always guarantees the strong heredity.

## 3 Simulation Studies

In this section, we compared the performance of the proposed hierarchical standardization methods to the corresponding methods using regular standardization. Robustness checks are also conducted in this section.

### 3.1 Simulation Setup

We considered nine simulation settings. In each setting, the training, validation and test datasets had sample sizes of 200, 200 and 10,000, respectively. All datasets were generated from Model (2) with $p = 10$: $Y = \eta + \sum_{j=1}^{10} \alpha_j X_j + \sum_{j=1}^{10} \beta_j X_j^2 + \sum_{1 \leq j < k \leq 10} \gamma_{jk} X_j X_k + \varepsilon$, where $\varepsilon \sim N(0, \sigma^2)$ and all $X_j$'s were generated independently from the standard normal distribution.

For variable selection, we chose some variables out of the 10 main effects to be *important*, i.e., they have non-zero coefficients. Other variables out of the 10 main effects have zero coefficients. To satisfy the heredity constraint, the corresponding two-factor interactions (2fi's) and quadratic effects generated from that subset were also set to be important. The details of the simulation setup are summarized in Table 1. Specifically, we include in Table 1 the signal-to-noise ratio (SNR) that has been used in the variable selection literature (e.g., [26, 33]). For a linear regression model: $Y = X\beta + \epsilon$, the SNR is defined to be $Var(X\beta)/Var(\epsilon)$, where $X$ is considered to be random. The SNR reflects how strong the signal is in the data compared to the noise. If the SNR is large, then the signal in the data is considered to be strong (compared to the noise) and the variable selection problem is considered to be easy. If the SNR is small, then the signal in the data is considered to be weak (compared to the noise) and the variable selection problem is considered to be difficult.

We compared five methods in our simulation:

1. Traditional lasso: Lasso under the regular standardization
2. Our lasso: Lasso under the hierarchical standardization
3. Traditional stepwise: Stepwise under the regular standardization
4. Our stepwise: Stepwise under the hierarchical standardization
5. SHIM: Strong Heredity Interaction Model based on Choi et al. [4]

All five methods were subjected to the same training, validation and test sets. For the traditional lasso, we applied the R package and function `glmnet` [9] to the generated training set. We first used default settings of `glmnet` for the regular standardization method [Eqs. (3) and (4)]. The `glmnet` function returned coefficients in their original scales. Then, for our lasso, we standardized the training set following Eq. (6) and then applied the default `glmnet` to the hierarchically standardized set. The `glmnet` function returned the coefficients, which were further transformed back based on Eq. (8). We used an example in Setting 1 to illustrate the hierarchical standardization. Please refer to "Appendix 3" for the detailed process of the hierarchical standardization.

**Table 1** Setup details for the simulations

| Settings | 1 | 2 | 3 | 4 | 5 | 6 | 7 | 8 | 9 |
|---|---|---|---|---|---|---|---|---|---|
| # of non-zero main effects | 3 | 3 | 3 | 4 | 4 | 4 | 5 | 5 | 5 |
| # of non-zero 2fi's | 3 | 3 | 3 | 6 | 6 | 6 | 10 | 10 | 10 |
| # of non-zero quadratic effects | 3 | 3 | 3 | 4 | 4 | 4 | 5 | 5 | 5 |
| Coef of non-zero main effects | 1 | 1 | 1 | 1 | 1 | 1 | 1 | 1 | 1 |
| Coef of non-zero 2fi's | 1 | 2 | 3 | 1 | 2 | 3 | 1 | 2 | 3 |
| Coef of non-zero quadratic effects | 1 | 2 | 3 | 1 | 2 | 3 | 1 | 2 | 3 |
| $\sigma$ for the error term | 8 | 16 | 15 | 8 | 16 | 15 | 8 | 16 | 15 |
| Signal-to-noise ratio | 0.19 | 0.15 | 0.37 | 0.28 | 0.23 | 0.58 | 0.39 | 0.33 | 0.83 |

The traditional and our stepwise methods followed a similar process, except that the validation of the stepwise methods did not need to be set to tune parameters. The R function `step` was used for the stepwise methods, with the full model to be the starting model. We used AIC as the criterion to select the best model. The mode of stepwise search involved both forward and backward. For the SHIM method, we adapted [4]'s R code and searched on a large $(\lambda_\gamma, \lambda_\beta)$ grid.

The traditional lasso, our lasso and SHIM required parameter tuning. For each parameter or combination of parameters, we fit the model and used the validation set to obtain the optimal value or combination of parameters with mean squared error (MSE) as the criterion. The model with the smallest MSE was selected. There were 50 replicates for each simulation setting.

### 3.2 The Strong Heredity Constraint

One of the most attractive features of the hierarchical standardization is that it automatically maintains heredity. In this part, we compare the performance of the five methods on maintaining the strong heredity using the following measurement based on the selected model:

$$\text{Maintenance of the Strong Heredity (MSH)} = \frac{\text{\# of correctly identified non-zero parent main effects}}{\text{Theoretical total \# of non-zero parent main effects}} \quad (10)$$

In Table 2, we summarize the average MSH measurement for the traditional lasso and traditional stepwise. The values of MSH varied from 0.273 to 0.54 for the traditional lasso and from 0.238 to 0.425 for the traditional stepwise. Thus, a substantial loss in heredity was noted in all nine settings for the two traditional methods. For instance, in Setting 1, the traditional lasso had an average MSH = 0.383, implying that 61.7% of the main effects that should have been important failed to be selected by the traditional lasso when using regular standardization. In comparison, our lasso and our stepwise methods based on the hierarchical standardization had MSH = 100% (SHIM also had MSH = 100%). The example in "Appendix 3" illustrates how the hierarchical standardization yields a 100% MSH as a guarantee of the strong heredity.

### 3.3 Variable Selection Performance and Prediction Accuracy

We now compare the five methods in terms of their variable selection performance and prediction accuracy. For the variable selection performance, we considered two common measures of *sensitivity* and *specificity*. For both measures, a higher value indicates better variable selection performance.

Table 2 The maintenance of strong heredity (MSH) measurement for traditional lasso and traditional stepwise

| Settings | | 1 | 2 | 3 | 4 | 5 | 6 | 7 | 8 | 9 |
|---|---|---|---|---|---|---|---|---|---|---|
| Traditional lasso | Mean | 0.383 | 0.273 | 0.321 | 0.45 | 0.299 | 0.4 | 0.54 | 0.386 | 0.498 |
| | s.e. | (0.025) | (0.032) | (0.025) | (0.028) | (0.026) | (0.022) | (0.026) | (0.028) | (0.025) |
| Traditional stepwise | Mean | 0.335 | 0.238 | 0.267 | 0.414 | 0.274 | 0.3 | 0.425 | 0.313 | 0.307 |
| | s.e. | (0.023) | (0.02) | (0.017) | (0.022) | (0.024) | (0.019) | (0.025) | (0.021) | (0.019) |

$$\text{Sensitivity} = \frac{\text{\# of correctly identified important variables in the selected model}}{\text{\# of important variables in the true model}}$$

$$\text{Specificity} = \frac{\text{\# of correctly identified unimportant variables in the selected model}}{\text{\# of unimportant variables in the true model}}$$

Figure 1 compares the distributions of sensitivity values for five methods across nine settings. It is apparent that our lasso and our stepwise outperform the traditional lasso and traditional stepwise, respectively. Substantial gain in sensitivity is often observed using the hierarchical standardization. For instance, in Setting 2, our lasso had a sensitivity of 72.2% (a median of 77.8%). In comparison, the traditional lasso had a sensitivity of 55.6% (a median of 55.6%) (all values are summarized in Table 3). Note that the nine settings in Table 1 constitute a $3 \times 3$ full factorial design in terms of the number of important factors (varying from 3 to 5 and corresponding to the three panels in each column) and the ratio between the 2fi coefficient and the corresponding main effect coefficient (varying from 1 to 3 and corresponding to the three panels in each row). As shown in the three panels of each row in Figure 1, the gain in the sensitivity from the hierarchical standardization compared to the corresponding traditional method becomes larger when the ratio of the 2fi coefficient over the parent main effect coefficient increases. In comparison, when the number of important factors increases from 3 to 5, as shown in the three panels in each column, the gain appears to be relatively stable.

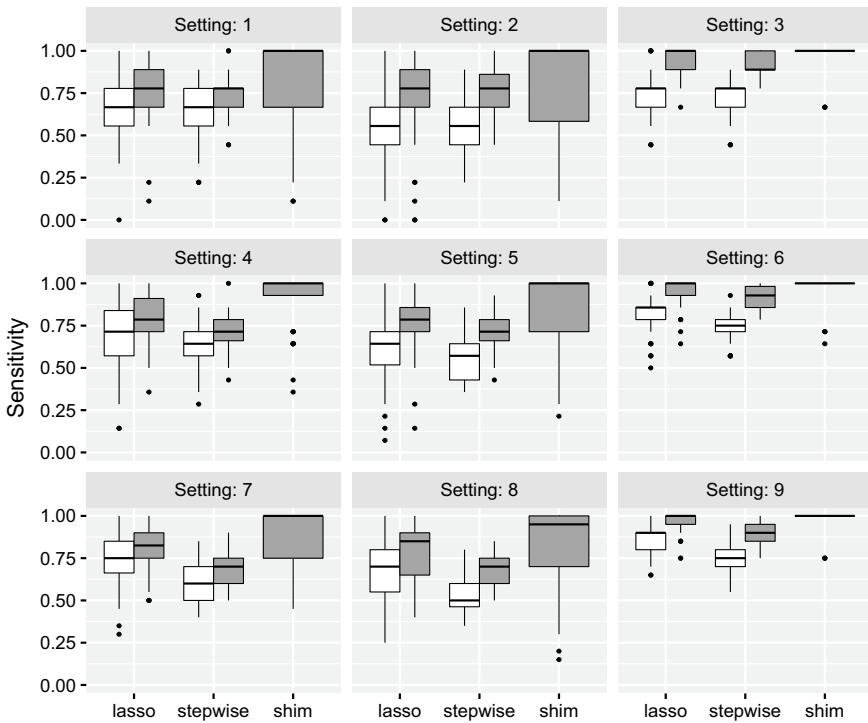

**Fig. 1** Boxplots of the sensitivity of our (grey boxes) and traditional (white boxes) methods and the SHIM. *Note 1* We compare all five methods within each setting

Figure 2 compares the distributions of specificity values of five methods across the nine settings. Because the hierarchical standardization forces parent main effects to be selected when a 2fi is in the model, our lasso and our stepwise had, not surprisingly, smaller specificity values compared to the corresponding traditional lasso and traditional stepwise. However, a comparison between Figs. 1 and 2 shows that the gain in sensitivity by using our methods generally outweighs the loss in specificity across all nine settings.

The mean, median and standard error values for both sensitivity and specificity of each method across all nine settings are given in Table 3. We further investigate whether the decrease in sensitivity is mainly due to the inclusion of many main effects, by breaking down the sensitivity and specificity into main effects, 2fi's, and quadratic effects in Table 4 (standard errors are in parentheses). It can be seen that methods with hierarchical standardization indeed include more unimportant main

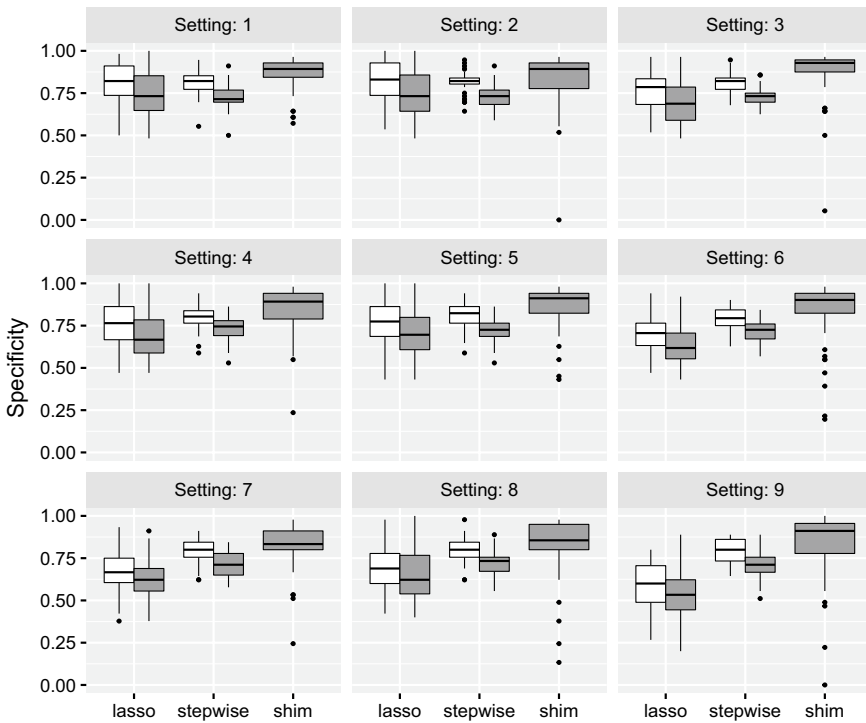

**Fig. 2** Boxplots of the specificity of our (grey boxes) and traditional (white boxes) methods and the SHIM. *Note 2* We compare all five methods within each setting

Table 3  Sensitivity and specificity for our and traditional methods and the SHIM

| Settings | | 1 | 2 | 3 | 4 | 5 | 6 | 7 | 8 | 9 |
|---|---|---|---|---|---|---|---|---|---|---|
| Sensitivity | | | | | | | | | | |
| Our lasso | Mean | 0.758 | 0.722 | 0.947 | 0.784 | 0.764 | 0.95 | 0.802 | 0.784 | 0.957 |
| | Median | 0.778 | 0.778 | 1 | 0.786 | 0.786 | 1 | 0.825 | 0.85 | 1 |
| | s.e. | (0.026) | (0.036) | (0.011) | (0.022) | (0.025) | (0.012) | (0.019) | (0.022) | (0.009) |
| Traditional lasso | Mean | 0.671 | 0.556 | 0.76 | 0.681 | 0.606 | 0.814 | 0.739 | 0.661 | 0.862 |
| | Median | 0.667 | 0.556 | 0.778 | 0.714 | 0.643 | 0.857 | 0.75 | 0.7 | 0.9 |
| | s.e. | (0.03) | (0.033) | (0.019) | (0.03) | (0.029) | (0.016) | (0.023) | (0.026) | (0.012) |
| Our stepwise | Mean | 0.747 | 0.753 | 0.92 | 0.726 | 0.73 | 0.91 | 0.69 | 0.685 | 0.895 |
| | Median | 0.778 | 0.778 | 0.889 | 0.714 | 0.714 | 0.929 | 0.7 | 0.7 | 0.9 |
| | s.e. | (0.018) | (0.018) | (0.012) | (0.015) | (0.015) | (0.01) | (0.015) | (0.013) | (0.01) |
| Traditional stepwise | Mean | 0.624 | 0.556 | 0.729 | 0.637 | 0.549 | 0.744 | 0.593 | 0.539 | 0.744 |
| | Median | 0.667 | 0.556 | 0.778 | 0.643 | 0.571 | 0.75 | 0.6 | 0.5 | 0.75 |
| | s.e. | (0.024) | (0.023) | (0.017) | (0.019) | (0.017) | (0.013) | (0.018) | (0.015) | (0.013) |
| SHIM | Mean | 0.849 | 0.807 | 0.98 | 0.913 | 0.834 | 0.976 | 0.897 | 0.796 | 0.985 |
| | Median | 1 | 1 | 1 | 1 | 1 | 1 | 1 | 0.95 | 1 |
| | s.e. | (0.038) | (0.037) | (0.011) | (0.023) | (0.033) | (0.012) | (0.023) | (0.035) | (0.008) |
| Specificity | | | | | | | | | | |
| Our lasso | Mean | 0.736 | 0.754 | 0.689 | 0.691 | 0.71 | 0.631 | 0.622 | 0.644 | 0.536 |
| | Median | 0.732 | 0.732 | 0.688 | 0.667 | 0.696 | 0.618 | 0.622 | 0.622 | 0.533 |
| | s.e. | (0.018) | (0.021) | (0.016) | (0.02) | (0.019) | (0.017) | (0.018) | (0.02) | (0.018) |
| Traditional lasso | Mean | 0.801 | 0.821 | 0.768 | 0.757 | 0.773 | 0.7 | 0.682 | 0.705 | 0.596 |
| | Median | 0.821 | 0.830 | 0.786 | 0.765 | 0.775 | 0.706 | 0.667 | 0.689 | 0.6 |
| | s.e. | (0.018) | (0.018) | (0.015) | (0.019) | (0.018) | (0.016) | (0.019) | (0.02) | (0.017) |

**Table 3** (continued)

| Settings | | 1 | 2 | 3 | 4 | 5 | 6 | 7 | 8 | 9 |
|---|---|---|---|---|---|---|---|---|---|---|
| Our stepwise | Mean | 0.727 | 0.728 | 0.728 | 0.73 | 0.725 | 0.722 | 0.712 | 0.72 | 0.712 |
| | Median | 0.714 | 0.732 | 0.732 | 0.745 | 0.725 | 0.725 | 0.711 | 0.733 | 0.711 |
| | s.e. | (0.009) | (0.009) | (0.007) | (0.01) | (0.01) | (0.009) | (0.01) | (0.01) | (0.011) |
| Traditional stepwise | Mean | 0.808 | 0.816 | 0.814 | 0.797 | 0.808 | 0.791 | 0.793 | 0.799 | 0.791 |
| | Median | 0.821 | 0.821 | 0.821 | 0.804 | 0.824 | 0.794 | 0.8 | 0.8 | 0.8 |
| | s.e. | (0.009) | (0.009) | (0.008) | (0.009) | (0.01) | (0.009) | (0.009) | (0.01) | (0.01) |
| SHIM | Mean | 0.866 | 0.838 | 0.867 | 0.836 | 0.859 | 0.829 | 0.83 | 0.815 | 0.809 |
| | Median | 0.893 | 0.893 | 0.929 | 0.892 | 0.912 | 0.902 | 0.833 | 0.856 | 0.911 |
| | s.e. | (0.015) | (0.024) | (0.022) | (0.021) | (0.019) | (0.027) | (0.019) | (0.026) | (0.028) |

**Table 4** Sensitivity and specificity breakdown by ME/2FI/QE

| | Setting 1 | | | | Setting 2 | | | | Setting 3 | | | |
|---|---|---|---|---|---|---|---|---|---|---|---|---|
| | Lasso | | Stepwise | | Lasso | | Stepwise | | Lasso | | Stepwise | |
| | Our | Trad | Our | Trad | Our | Trad | Our | Trad | Our | Trad | Our | Trad |
| *Sensi* | | | | | | | | | | | | |
| ME | 0.96 (0.021) | 0.66 (0.044) | 0.987 (0.009) | 0.64 (0.041) | 0.9 (0.036) | 0.347 (0.044) | 0.993 (0.007) | 0.373 (0.039) | 1 (0.000) | 0.433 (0.041) | 1 (0.000) | 0.393 (0.034) |
| 2FI | 0.5 (0.045) | 0.533 (0.046) | 0.473 (0.039) | 0.473 (0.038) | 0.493 (0.048) | 0.513 (0.046) | 0.487 (0.040) | 0.5 (0.040) | 0.853 (0.032) | 0.86 (0.030) | 0.807 (0.030) | 0.84 (0.029) |
| QE | 0.813 (0.036) | 0.82 (0.036) | 0.78 (0.035) | 0.76 (0.038) | 0.773 (0.043) | 0.807 (0.041) | 0.78 (0.036) | 0.793 (0.033) | 0.987 (0.009) | 0.987 (0.009) | 0.953 (0.017) | 0.953 (0.017) |
| *Speci* | | | | | | | | | | | | |
| ME | 0.203 (0.037) | 0.771 (0.026) | 0.109 (0.021) | 0.809 (0.021) | 0.26 (0.047) | 0.8 (0.026) | 0.114 (0.020) | 0.814 (0.022) | 0.12 (0.026) | 0.743 (0.027) | 0.083 (0.016) | 0.797 (0.022) |
| 2FI | 0.814 (0.018) | 0.808 (0.019) | 0.815 (0.010) | 0.806 (0.010) | 0.827 (0.019) | 0.823 (0.019) | 0.815 (0.009) | 0.814 (0.009) | 0.774 (0.017) | 0.772 (0.016) | 0.82 (0.007) | 0.815 (0.008) |
| QE | 0.803 (0.021) | 0.794 (0.022) | 0.82 (0.019) | 0.823 (0.020) | 0.811 (0.024) | 0.826 (0.022) | 0.817 (0.020) | 0.829 (0.019) | 0.749 (0.022) | 0.766 (0.021) | 0.82 (0.019) | 0.823 (0.019) |
| | Setting 4 | | | | Setting 5 | | | | Setting 6 | | | |
| | Lasso | | Stepwise | | Lasso | | Stepwise | | Lasso | | Stepwise | |
| | Our | Trad | Our | Trad | Our | Trad | Our | Trad | Our | Trad | Our | Trad |
| *Sensi* | | | | | | | | | | | | |
| ME | 0.99 (0.007) | 0.67 (0.044) | 1 (0.000) | 0.65 (0.034) | 0.965 (0.019) | 0.405 (0.040) | 1 (0.000) | 0.38 (0.037) | 1 (0.000) | 0.515 (0.033) | 1 (0.000) | 0.425 (0.027) |

**Table 4** (continued)

|  | Setting 4 | | | | | | Setting 5 | | | | | | Setting 6 | | | | | |
|---|---|---|---|---|---|---|---|---|---|---|---|---|---|---|---|---|---|---|
|  | Lasso | | Stepwise | | Trad | | Lasso | | Stepwise | | Trad | | Lasso | | Stepwise | | Trad | |
|  | Our | Trad | Our | Trad | Our | Trad | Our | Trad | Our | Trad | Our | Trad | Our | Trad | Our | Trad | Our | Trad |
| 2FI | 0.627 (0.037) | 0.603 (0.037) | 0.523 (0.026) | 0.53 (0.030) | | | 0.59 (0.038) | 0.597 (0.038) | 0.533 (0.029) | 0.52 (0.028) | | | 0.89 (0.026) | 0.9 (0.021) | 0.82 (0.021) | 0.817 (0.021) | | |
| QE | 0.815 (0.029) | 0.81 (0.031) | 0.755 (0.028) | 0.785 (0.025) | | | 0.825 (0.028) | 0.82 (0.029) | 0.755 (0.028) | 0.76 (0.029) | | | 0.99 (0.007) | 0.985 (0.008) | 0.955 (0.014) | 0.955 (0.014) | | |

*Speci*

|  | Lasso | | Stepwise | | Trad | | Lasso | | Stepwise | | Trad | | Lasso | | Stepwise | | Trad | |
|---|---|---|---|---|---|---|---|---|---|---|---|---|---|---|---|---|---|---|
| ME | 0.147 (0.036) | 0.733 (0.029) | 0.117 (0.019) | 0.763 (0.023) | | | 0.157 (0.038) | 0.77 (0.024) | 0.1 (0.021) | 0.807 (0.022) | | | 0.053 (0.020) | 0.677 (0.027) | 0.083 (0.018) | 0.777 (0.023) | | |
| 2FI | 0.766 (0.018) | 0.761 (0.018) | 0.812 (0.010) | 0.801 (0.010) | | | 0.785 (0.018) | 0.773 (0.019) | 0.808 (0.010) | 0.807 (0.010) | | | 0.712 (0.017) | 0.705 (0.017) | 0.81 (0.009) | 0.795 (0.009) | | |
| QE | 0.75 (0.028) | 0.753 (0.029) | 0.81 (0.022) | 0.807 (0.024) | | | 0.78 (0.026) | 0.777 (0.024) | 0.81 (0.023) | 0.817 (0.024) | | | 0.687 (0.026) | 0.69 (0.027) | 0.787 (0.023) | 0.78 (0.025) | | |

|  | Setting 7 | | | | | | Setting 8 | | | | | | Setting 9 | | | | | |
|---|---|---|---|---|---|---|---|---|---|---|---|---|---|---|---|---|---|---|
|  | Lasso | | Stepwise | | | | Lasso | | Stepwise | | | | Lasso | | Stepwise | | | |
|  | Our | Trad | Our | Trad | | | Our | Trad | Our | Trad | | | Our | Trad | Our | Trad | | |

*Sensi*

| ME | 1 (0.000) | 0.72 (0.036) | 1 (0.000) | 0.6 (0.033) | | | 0.992 (0.006) | 0.472 (0.037) | 1 (0.000) | 0.38 (0.031) | | | 1 (0.000) | 0.596 (0.032) | 1 (0.000) | 0.392 (0.028) | | |
| 2FI | 0.674 (0.030) | 0.68 (0.027) | 0.52 (0.023) | 0.526 (0.023) | | | 0.648 (0.034) | 0.66 (0.031) | 0.52 (0.020) | 0.524 (0.021) | | | 0.914 (0.018) | 0.926 (0.016) | 0.808 (0.018) | 0.808 (0.019) | | |
| QE | 0.86 (0.026) | 0.876 (0.025) | 0.72 (0.030) | 0.72 (0.031) | | | 0.848 (0.027) | 0.852 (0.027) | 0.7 (0.028) | 0.728 (0.030) | | | 1 (0.000) | 1 (0.000) | 0.964 (0.012) | 0.968 (0.010) | | |

**Table 4** (continued)

| | Setting 7 | | | | | | Setting 8 | | | | | | Setting 9 | | | | | |
|---|---|---|---|---|---|---|---|---|---|---|---|---|---|---|---|---|---|---|
| | Lasso | | Stepwise | | | | Lasso | | Stepwise | | | | Lasso | | Stepwise | | | |
| | Our | Trad | Our | Trad | | | Our | Trad | Our | Trad | | | Our | Trad | Our | Trad | | |
| *Speci* | | | | | | | | | | | | | | | | | | |
| ME | 0.04 | 0.648 | 0.08 | 0.772 | | | 0.092 | 0.7 | 0.096 | 0.76 | | | 0.012 | 0.592 | 0.088 | 0.784 | | |
| | (0.017) | (0.028) | (0.017) | (0.026) | | | (0.029) | (0.029) | (0.017) | (0.025) | | | (0.009) | (0.030) | (0.017) | (0.025) | | |
| 2FI | 0.696 | 0.689 | 0.791 | 0.796 | | | 0.715 | 0.708 | 0.796 | 0.803 | | | 0.6 | 0.598 | 0.789 | 0.79 | | |
| | (0.019) | (0.020) | (0.010) | (0.010) | | | (0.021) | (0.021) | (0.010) | (0.010) | | | (0.020) | (0.018) | (0.011) | (0.009) | | |
| QE | 0.684 | 0.668 | 0.788 | 0.792 | | | 0.7 | 0.692 | 0.816 | 0.808 | | | 0.608 | 0.588 | 0.804 | 0.8 | | |
| | (0.034) | (0.032) | (0.029) | (0.029) | | | (0.033) | (0.031) | (0.026) | (0.030) | | | (0.033) | (0.032) | (0.028) | (0.027) | | |

effects compared to the methods with traditional standardization, while the inclusion of unimportant 2fi's and quadratic effects is comparable between the two types of methods. The inclusion of more unimportant ME's can be considered as a price we pay for the satisfaction of the strong heredity.

Table 5 shows the prediction accuracy of the five methods in terms of the mean squared error (MSE) criteria. The results indicate that the proposed methods and the corresponding traditional methods have comparable mean (and median) MSE values. SHIM performed very well in all settings we considered. This is not surprising, as SHIM was developed particularly for models following heredity. Note that the goal of our work is to propose a general procedure to help any variable selection method achieve strong heredity in the selected model. The purpose is not to show that our lasso is better compared to some other variable selection methods, like SHIM. Our aim is to demonstrate that using the hierarchical standardization, our lasso can guarantee strong heredity in the selected model and can outperform the traditional lasso in certain aspects, as illustrated in the simulation studies. Comparing our lasso with SHIM, the performance of our lasso depends on the "intrinsic" variable selection ability of the lasso. It is not surprising that a non-convex penalty function like SHIM has better finite sample performance compared to a convex penalty, like the lasso [8, 23].

**Table 5** Prediction MSE's for our and traditional methods and the SHIM using the test set

| Settings | 1 | 2 | 3 | 4 | 5 | 6 | 7 | 8 | 9 |
|---|---|---|---|---|---|---|---|---|---|
| *Our lasso* | | | | | | | | | |
| Mean | 72.124 | 284.622 | 256.798 | 75.619 | 297.387 | 271.752 | 79.282 | 311.603 | 287.908 |
| Median | 71.994 | 283.027 | 257.010 | 75.661 | 296.466 | 269.606 | 79.601 | 314.080 | 286.936 |
| s.e. | (0.35) | (1.322) | (1.364) | (0.404) | (1.49) | (2.05) | (0.433) | (1.577) | (2.215) |
| *Traditional lasso* | | | | | | | | | |
| Mean | 72.031 | 283.747 | 255.85 | 75.548 | 296.544 | 270.153 | 79.145 | 310.786 | 286.12 |
| Median | 72.129 | 282.290 | 255.833 | 75.380 | 295.658 | 267.733 | 79.416 | 312.253 | 285.067 |
| s.e. | (0.339) | (1.293) | (1.309) | (0.395) | (1.436) | (1.874) | (0.437) | (1.563) | (2.14) |
| *Our stepwise* | | | | | | | | | |
| Mean | 87.601 | 348.352 | 304.474 | 90.566 | 358.771 | 315.791 | 96.658 | 381.064 | 332.744 |
| Median | 86.916 | 344.507 | 303.415 | 88.933 | 353.682 | 311.739 | 96.074 | 378.790 | 331.037 |
| s.e. | (1.045) | (4.147) | (3.779) | (1.1) | (4.32) | (3.958) | (1.177) | (4.626) | (5.161) |
| *Traditional stepwise* | | | | | | | | | |
| Mean | 87.725 | 346.507 | 303.572 | 90.759 | 359.472 | 318.913 | 96.14 | 376.628 | 329.51 |
| Median | 86.629 | 343.262 | 302.090 | 88.977 | 355.238 | 313.346 | 96.186 | 376.269 | 328.793 |
| s.e. | (1.067) | (4.092) | (3.749) | (0.956) | (4.405) | (3.914) | (1.136) | (4.562) | (4.546) |
| *SHIM* | | | | | | | | | |
| Mean | 69.653 | 281.951 | 247.938 | 71.775 | 292.163 | 258.877 | 75.004 | 308.916 | 274.383 |
| Median | 68.801 | 280.712 | 245.219 | 71.363 | 288.710 | 253.226 | 75.071 | 304.416 | 270.845 |
| s.e. | (0.48) | (1.872) | (1.879) | (0.514) | (2.245) | (2.541) | (0.594) | (3.035) | (3.234) |

### 3.4 Robustness Check

#### 3.4.1 True Model Violating Strong Heredity

We conducted a robustness check to assess the performance of our heredity-enforced methods when the true model did not satisfy the strong heredity constraint. For each of the nine settings in Table 1, we studied the performance of our lasso and our stepwise by assuming that the coefficients of some of the important parent main effects in the original model now equal zero. We reduced the number of important main effects from 3 (for Settings 1–3), 4 (for Settings 4–6) and 5 (for Settings 7–9) to 2. Consequently, the resulting models no longer followed heredity. The reduced models are denoted by "R" in Table 6 while the original (full) models are denoted by "F". We defined the loss in mean sensitivity (loss in mean specificity and loss in mean MSE are defined in a similar way) as the ratio of the difference between sensitivity (or specificity, or MSE) of models F and R over sensitivity (or specificity, or MSE) of model F. Note that the "loss" can take a positive value when the reduced model R results in better sensitivity and specificity compared to the original model F.

Table 6 shows that both hierarchical standardization-based lasso and stepwise methods demonstrate robustness in terms of all three criteria: sensitivity, specificity and MSE. For mean sensitivity, the loss ranges from 0.74% (Setting 3) to 6.6% (Setting 1) for our lasso and from 0% (Setting 3) to 7.88% (Setting 8) for our stepwise. For mean specificity, the loss ranges from 0.27% (Setting 1) to 5.97% (Setting 9) for our lasso and from 0.83% (Setting 1) to 5.97% (Setting 8) for our stepwise. The loss in mean MSE ranges from 0.24% (Setting 3) to 1.66% (Setting 7) for our lasso and from 0.28% (Setting 3) to 2.38% (Setting 7) for our stepwise. Overall, the proposed hierarchical standardization methods are fairly robust against violation of the heredity assumption.

#### 3.4.2 Skewed Variable Distribution

We conduct another robustness check to compare between different location-scale transformations on the main effects in the hierarchical standardization. As pointed out in Sect. 2, there are many different ways to standardize the main effects for the first step of hierarchical standardization. While the mean-standard deviation transformation is most widely used, standardization using other robust measures of location and spread such as the median-IQR transformation is also possible. This section investigates the performance of hierarchical standardization using the median-IQR transformation.

In addition to the normal distribution used in the main analysis, we also let $X$'s have log-normal distributions which are skewed. The mean-standard deviation transformation and the median-IQR transformation are applied to normally and log-normally distributed $X$'s for comparison. The simulation setup is summarized in Table 7. Setting 1 comes from the main analysis and serves as a benchmark.

As expected, the hierarchical standardization can guarantee the strong heredity. With the median-IQR transformation in the robustness check, the maintenance of the strong heredity (MSH) is 100% for our lasso and our stepwise. We report the mean sensitivity, mean specificity and mean MSE in Table 8. When we have skewed or heavy-tailed $X$'s (such as the log-normal distribution), the mean-SD transformation is not a very good transformation in the hierarchical standardization. Comparing R1 to R3, when $X$'s are log-normally distributed, using median-IQR in the hierarchical standardization generates much lower MSE than using mean-SD. The sensitivity and specificity are still comparable between traditional and our variable selection methods. However, when $X$'s have a well-behaved distribution (Settings 1 and R2), median-IQR does not generate much different performance from the mean-SD transformation.

**Table 6** Mean sensitivity, mean specificity and mean MSE for our and traditional methods in the robustness check

| Settings | 1 | 2 | 3 | 4 | 5 | 6 | 7 | 8 | 9 |
|---|---|---|---|---|---|---|---|---|---|
| *Mean sensitivity* | | | | | | | | | |
| Our lasso (R) | 0.708 | 0.7 | 0.94 | 0.74 | 0.718 | 0.933 | 0.759 | 0.74 | 0.946 |
| Our lasso (F) | 0.758 | 0.722 | 0.947 | 0.784 | 0.764 | 0.95 | 0.802 | 0.784 | 0.957 |
| Traditional lasso (R) | 0.66 | 0.58 | 0.79 | 0.72 | 0.643 | 0.875 | 0.734 | 0.698 | 0.906 |
| Our stepwise (R) | 0.723 | 0.715 | 0.92 | 0.687 | 0.687 | 0.893 | 0.639 | 0.631 | 0.873 |
| Our stepwise (F) | 0.747 | 0.753 | 0.92 | 0.726 | 0.73 | 0.91 | 0.69 | 0.685 | 0.895 |
| Traditional stepwise (R) | 0.623 | 0.56 | 0.772 | 0.623 | 0.588 | 0.797 | 0.581 | 0.551 | 0.787 |
| *Mean specificity* | | | | | | | | | |
| Our lasso (R) | 0.738 | 0.743 | 0.674 | 0.667 | 0.686 | 0.611 | 0.595 | 0.609 | 0.504 |
| Our lasso (F) | 0.736 | 0.754 | 0.689 | 0.691 | 0.71 | 0.631 | 0.622 | 0.644 | 0.536 |
| Traditional lasso (R) | 0.812 | 0.823 | 0.765 | 0.749 | 0.765 | 0.691 | 0.695 | 0.708 | 0.602 |
| Our stepwise (R) | 0.721 | 0.717 | 0.705 | 0.695 | 0.693 | 0.694 | 0.679 | 0.677 | 0.672 |
| Our stepwise (F) | 0.727 | 0.728 | 0.728 | 0.73 | 0.725 | 0.722 | 0.712 | 0.72 | 0.712 |
| Traditional stepwise (R) | 0.813 | 0.814 | 0.818 | 0.801 | 0.806 | 0.797 | 0.797 | 0.803 | 0.797 |
| *Mean MSE* | | | | | | | | | |
| Our lasso (R) | 71.692 | 283.799 | 256.18 | 74.428 | 296.105 | 270.41 | 77.967 | 310.507 | 286.401 |
| Our lasso (F) | 72.124 | 284.622 | 256.798 | 75.619 | 297.387 | 271.752 | 79.282 | 311.603 | 287.908 |
| Traditional lasso (R) | 71.426 | 282.819 | 255.261 | 74.288 | 295.283 | 269.111 | 77.77 | 308.721 | 284.62 |
| Our stepwise (R) | 86.91 | 347.103 | 305.313 | 89.85 | 360.112 | 312.829 | 94.362 | 377.414 | 329.044 |
| Our stepwise (F) | 87.601 | 348.352 | 304.474 | 90.566 | 358.771 | 315.791 | 96.658 | 381.064 | 332.744 |
| Traditional stepwise (R) | 86.948 | 345.811 | 301.02 | 90.306 | 357.931 | 316.113 | 94.359 | 374.717 | 329.263 |

**Table 7** Setup for robustness check on main-effect transformation

| Settings | 1 | R1 | R2 | R3 |
|---|---|---|---|---|
| # of main effects | 10 | 10 | 10 | 10 |
| # of non-zero main effects | 3 | 3 | 3 | 3 |
| # of non-zero 2fi's | 3 | 3 | 3 | 3 |
| # of non-zero quadratic effects | 3 | 3 | 3 | 3 |
| Coef of non-zero main effects | 1 | 1 | 1 | 1 |
| Coef of non-zero 2fi's | 1 | 1 | 1 | 1 |
| Coef of non-zero quadratic effects | 1 | 1 | 1 | 1 |
| Distribution of $X$'s | $N(0, 1^2)$ | lognormal$(0, 1^2)$ | $N(0, 1^2)$ | lognormal$(0, 1^2)$ |
| Transformation of main effects | Mean-SD | Mean-SD | Median-IQR | Median-IQR |
| $\sigma$ for the error term | 8 | 8 | 8 | 8 |
| Signal-to-noise ratio | 0.188 | 364.018 | 0.188 | 364.018 |

**Table 8** Mean sensitivity, mean specificity and mean MSE for main-effect transformation comparison in the robustness check

| Settings | | 1 | R1 | R2 | R3 |
|---|---|---|---|---|---|
| *Mean sensitivity* | | | | | |
| Our lasso | Mean | 0.758 | 0.956 | 0.773 | 0.987 |
| | s.e. | 0.026 | 0.011 | 0.023 | 0.005 |
| Traditional lasso | Mean | 0.671 | 0.911 | 0.671 | 0.911 |
| | s.e. | 0.03 | 0.013 | 0.03 | 0.013 |
| Our stepwise | Mean | 0.747 | 0.989 | 0.749 | 0.987 |
| | s.e. | 0.018 | 0.005 | 0.018 | 0.006 |
| Traditional stepwise | Mean | 0.624 | 0.782 | 0.624 | 0.782 |
| | s.e. | 0.024 | 0.015 | 0.024 | 0.015 |
| *Mean specificity* | | | | | |
| Our lasso | Mean | 0.736 | 0.609 | 0.725 | 0.7 |
| | s.e. | 0.018 | 0.027 | 0.019 | 0.023 |
| Traditional lasso | Mean | 0.801 | 0.865 | 0.801 | 0.865 |
| | s.e. | 0.018 | 0.013 | 0.018 | 0.013 |
| Our stepwise | Mean | 0.728 | 0.758 | 0.728 | 0.761 |
| | s.e. | 0.009 | 0.009 | 0.01 | 0.011 |
| Traditional stepwise | Mean | 0.808 | 0.884 | 0.808 | 0.884 |
| | s.e. | 0.009 | 0.007 | 0.009 | 0.007 |
| *Mean MSE* | | | | | |
| Our lasso | Mean | 72.124 | 1215.682 | 72.184 | 668.339 |
| | s.e. | 0.35 | 1052.648 | 0.351 | 539.642 |
| Traditional lasso | Mean | 72.031 | 359.858 | 72.031 | 359.858 |
| | s.e. | 0.339 | 263.66 | 0.339 | 263.66 |
| Our stepwise | Mean | 87.601 | 744.947 | 87.745 | 326.007 |
| | s.e. | 1.045 | 567.759 | 1.158 | 170.878 |
| Traditional stepwise | Mean | 87.725 | 281.329 | 87.725 | 281.329 |
| | s.e. | 1.067 | 124.355 | 1.067 | 124.355 |

### 3.4.3 Additional Robustness Checks

We consider two extensions of the simulation setup to make our results more generalizable. First, we enlarge the main effect coefficients to see if the hierarchical standardization still maintains 100% MSH and generates comparable performance to traditional methods. Second, we consider the cases where some variables enter as active main effects but do not appear in active higher-order terms. The setup and results are presented in "Appendix 4".

Results indicate that for both extensions, the hierarchical standardization maintains 100% strong heredity. For both extensions, our method has comparable performance metrics (sensitivity, specificity and MSE) to the traditional variable selection counterpart. Hence, the hierarchical standardization can guarantee the strong heredity without sacrificing estimation or prediction performance too much.

## 4 Real Data Analysis

In this section, we review our analysis of a gene expression dataset to demonstrate the applicability and effectiveness of our methods. The dataset was obtained from the Cancer Genome Atlas (TCGA) study on glioblastoma. Glioblastoma is the most common primary brain tumor in adults. Patients with newly diagnosed glioblastoma have a median survival of approximately 1 year, with poor responses to all therapeutic modalities [19].

In the biological and medical research in general, and in this analysis in particular, *epistatic* effects, i.e., the gene-gene interactions, have received large attention in genome-wide association studies [5, 15]. In these cases, including our example, heredity is a desirable property of the selected model. Therefore, we applied the proposed hierarchical standardization method to identify important genes and inter- or intra-gene interactions that are related to patients' survival.

Overall, 538 participants were enrolled in the study. For each subject, we obtained his/her survival time and 12,042 gene expression profiles. Since glioblastoma is severe cancer, the censoring rate of the dataset was low (about 15%). We simply removed censored subjects, which left us with 449 subjects in the dataset. We treated the log-survival time as a continuous outcome for the remaining subjects. Following [28], we ranked all genes by their variances and kept the top 100 genes. We performed the same analysis with the top 150 and 200 genes. Genes may be associated with the phenotype in a nonlinear fashion. In this case, the inclusion of the quadratic terms (and higher-order terms) can be helpful to detect important genes that are related to the phenotype. Hence, we included all pairwise interaction terms (2fi's) and quadratic terms corresponding to these genes. 100, 150 and 200 main effects of genes resulted in 5150, 11,475 and 20,300 total terms in the model, respectively.

We randomly split the data into training, validation and test sets with a 3:1:1 ratio. Since the dataset was high-dimensional (i.e., the number of predictors was higher than the number of observations), stepwise (traditional and our versions) could not start with the full model. If the null model becomes the starting model instead,

**Table 9** Maintenance of the strong heredity (MSH) for traditional lasso and traditional stepwise in the glioblastoma gene expression data analysis

|  | MSH (100 genes) | | MSH (150 genes) | | MSH (200 genes) | |
| --- | --- | --- | --- | --- | --- | --- |
|  | Mean (%) | s.e. | Mean (%) | s.e. | Mean (%) | s.e. |
| Traditional lasso | 6.39 | 0.028 | 6.68 | 0.034 | 10.39 | 0.043 |
| Traditional stepwise | 1.58 | 0.002 | 0.69 | 0.001 | 0.58 | 0.001 |

Traditional stepwise starts with the null model and stops when or before the number of predictors reach the number of observations

**Table 10** MSE of four methods in the glioblastoma gene expression data analysis

|  | MSE for 100 genes | | | MSE for 150 genes | | | MSE for 200 genes | | |
| --- | --- | --- | --- | --- | --- | --- | --- | --- | --- |
|  | Mean | Median | s.e. | Mean | Median | s.e. | Mean | Median | s.e. |
| Our lasso | 1.2362 | 1.2425 | 0.032 | 1.2215 | 1.2173 | 0.033 | 1.2236 | 1.2171 | 0.035 |
| Traditional lasso | 1.2376 | 1.1964 | 0.032 | 1.2334 | 1.2028 | 0.033 | 1.2414 | 1.2082 | 0.033 |
| Our stepwise | 2.1780 | 2.1305 | 0.059 | 2.2555 | 2.1626 | 0.071 | 2.2004 | 2.2386 | 0.061 |
| Traditional stepwise | 2.0310 | 1.9918 | 0.043 | 2.1003 | 2.0371 | 0.056 | 2.1533 | 2.0979 | 0.054 |

Our stepwise and traditional stepwise start with the null model and stop when or before the number of predictors reach the number of observations traditional and our stepwise could not proceed when the number of selected variables reached the number of observations. We employed traditional and our stepwise methods that start with the null model and report results in Table 10, for comparison purpose. In addition, we performed traditional lasso and our lasso, which were fit to the training set. The tuning parameters were selected using the validation set. We then calculated the mean squared error (MSE) using the test set. The entire procedure was repeated 50 times.

The analysis results indicate that the traditional lasso and traditional stepwise violate the heredity constraint when considering 100, 150 and 200 genes. The mean values and standard errors of maintenance of the strong heredity (MSH) for traditional lasso and traditional stepwise are reported in Table 9. Our lasso and our stepwise satisfy the heredity constraint in all three situations, as expected (i.e., MSH = 100%). The prediction performance in terms of MSE of the two methods is summarized in Table 10. The table reports the analysis conducted with the top 100 (left block), 150 (middle block) and 200 genes (right block), their two-way interactions, and their quadratic effects. The average MSE over 50 replicates and the standard errors are reported. Our lasso has smaller MSEs compared to the traditional lasso in all three situations. Our stepwise has larger but comparable MSEs compared to the traditional stepwise.

We have taken a further look into the selected genes. The lasso with our hierarchical standardization selects the following 26 genes out of the screened 200 genes.

| BBOX1  | CLK4   | SALL1 | SUHW1 | MYO9B   | C9orf45 | JOSD3  |
|--------|--------|-------|-------|---------|---------|--------|
| OSBPL8 | PIR    | EVC   | DMC1  | SLC12A4 | SLCO2B1 | DIMT1L |
| TERF2IP| ZNF212 | SELL  | WDR45 | PSCD3   | PYY     | SIRT6  |
| ABR    | GML    | GRB2  | EPB42 | FPGT    |         |        |

We have reviewed related literature in the cancer research to validate the selected genes. The gene with the largest absolute coefficient is "PYY". Its relationship to glioblastoma has been largely studied, for example, in [16, 17], and so on. The selected gene with the second largest absolute coefficient is "OSBPL8", which is also shown to affect the brain [24, 25]. This demonstrates that the lasso method with our proposed hierarchical standardization selects some biologically meaningful genes.

## 5 Conclusion

In this paper, we proposed the hierarchical standardization that enforces strong heredity for variable selection methods. The hierarchical standardization is applicable to any variable selection method and is easy to implement in practice. In addition to its capability to retain the heredity structure of the model, the simulation studies have shown that it also performs well in terms of commonly used criteria of sensitivity, specificity and MSE.

Compared to the structured variable selection methods that are designed especially for models following heredity, the proposed approach has several advantages. First, our method is more general and easily adaptive. On the contrary, structured variable selection methods are specially developed, and they may not be easily adapted to other penalty functions. For example, Yuan et al. [29]'s approach is an extension of the LARS algorithm, and it may not be easily extended to other popular methods, like SCAD [8].

Second, some structured variable selection methods achieve strong heredity at a cost. For the composite absolute penalties (CAP) family, both first-order and second-order terms are over-penalized to enforce the strong heredity. For example, in the two-way interaction model, each interaction term is penalized twice, and each main effect is penalized $p-1$ times. For the SHIM method, the interaction term with smaller main effects tends to be penalized more heavily to maintain the strong heredity in the selected model. These costs may prevent related methods from adequate model selection and estimation performances in certain scenarios. By contrast, our approach transforms the heredity-constrained variable selection problem to an unconstrained variable selection problem. An unconstrained variable selection method can yield the heredity-enforced selected model without any extra cost.

Third, the hierarchical standardization does not change the convexity of the variable selection method. Therefore, good statistical properties can be preserved, which means our approach can be numerically stable if the variable selection method is convex. On the contrary, although the SHIM method proposed by Choi et al. [4] generally performs well compared to competing methods, it has a non-convex objective function. As a common potential problem for all non-convex methods, the SHIM may have more than one local minimum instead of a unique global minimum. As a consequence of this, the solution to SHIM may depend on the starting value, and the performance of the method may not be stable, especially when the dimension of covariates is high. In addition, the SHIM may even fail to converge as the iteration may go back and forth between several local minima. In general, as a non-convex method, the SHIM may require additional computational efforts, which may prevent it from yielding adequate performance for some datasets.

**Acknowledgements** Sijian Wang gratefully acknowledges the support from *NIH 5 R01 HG007377-02*.

## Appendix 1: Proof of Eq. (8) and Guaranteed Strong Heredity

***Proof*** We expand Eq. (7) by replacing $\tilde{X}_j$ with $(X_j - \bar{X}_j)/s_j$:

$$\begin{aligned}
\hat{Y} =& \tilde{\eta}^H + \sum_{j=1}^{p} \tilde{\alpha}_j^H \frac{X_j - \bar{X}_j}{s_j} + \sum_{j=1}^{p} \tilde{\beta}_j^H \frac{(X_j - \bar{X}_j)^2}{s_j^2} \\
& + \sum_{1 \le j < k \le p} \tilde{\gamma}_{jk}^H \frac{(X_j - \bar{X}_j)(X_k - \bar{X}_k)}{s_j s_k} \\
=& \tilde{\eta}^H + \sum_{j=1}^{p} \tilde{\alpha}_j^H \frac{X_j}{s_j} + \sum_{j=1}^{p} \tilde{\beta}_j^H \frac{X_j^2 - 2\bar{X}_j X_j}{s_j^2} \\
& + \sum_{1 \le j < k \le p} \tilde{\gamma}_{jk}^H \frac{X_j X_k - \bar{X}_j X_k - \bar{X}_k X_j}{s_j s_k} \\
& - \sum_{j=1}^{p} \tilde{\alpha}_j^H \frac{\bar{X}_j}{s_j} + \sum_{j=1}^{p} \tilde{\beta}_j^H \frac{\bar{X}_j^2}{s_j^2} + \sum_{1 \le j < k \le p} \tilde{\gamma}_{jk}^H \frac{\bar{X}_j \bar{X}_k}{s_j s_k} \\
=& \sum_{j=1}^{p} \frac{\tilde{\alpha}_j^H}{s_j} X_j - 2 \sum_{j=1}^{p} \frac{\tilde{\beta}_j^H \bar{X}_j}{s_j^2} X_j - \sum_{1 \le j < k \le p} \frac{\tilde{\gamma}_{jk}^H}{s_j s_k}(\bar{X}_j X_k + \bar{X}_k X_j) \\
& + \sum_{j=1}^{p} \frac{\tilde{\beta}_j^H}{s_j^2} X_j^2 + \sum_{1 \le j < k \le p} \frac{\tilde{\gamma}_{jk}^H}{s_j s_k} X_j X_k + \text{Constant}
\end{aligned} \quad (11)$$

For a specific $j$ and a pair of $j$ and $k$, we can obtain from Eq. (11) that

$$\hat{\beta}_j^H = \frac{\tilde{\beta}_j^H}{s_j^2}, \quad \hat{\gamma}_{jk}^H = \frac{\tilde{\gamma}_{jk}^H}{s_j s_k} \quad (12)$$

For a specific $j$, the main effect of $X_j$ can be obtained as

$$\hat{\alpha}_j^H = \frac{\tilde{\alpha}_j^H}{s_j} - 2\frac{\bar{X}_j \tilde{\beta}_j^H}{s_j^2} - \sum_{\substack{k \neq j \\ 1 \leq k \leq p}} \frac{\bar{X}_k \tilde{\gamma}_{jk}^H}{s_j s_k} \qquad (13)$$

where the last term of Eq. (13) comes from re-arranging $\sum_{1 \leq j < k \leq p} \frac{\tilde{\gamma}_{jk}^H}{s_j s_k}(\bar{X}_j X_k + \bar{X}_k X_j)$. Hence, we have proven Eq. (8).

Next, we prove that, by applying Eq. (8) [equivalent to Eqs. (12) and (13)], the probability that the main effect [Eq. (13)] equals zero is zero when related higher-order terms are non-zero.

First, without loss of generality, if the quadratic term $X_j^2$ is selected while none of the 2-factor interactions related to $X_j$ is selected, then $\tilde{\beta}_j^H \neq 0$ and $\tilde{\gamma}_{jk}^H = 0$. Correspondingly,

$$\hat{\alpha}_j^H = \frac{\tilde{\alpha}_j^H}{s_j} - 2\frac{\bar{X}_j \tilde{\beta}_j^H}{s_j^2} \qquad (14)$$

If $X_j$ is not selected (i.e., the strong heredity is violated), then $\tilde{\alpha}_j^H = 0$. The proposed hierarchical standardization guarantees the strong heredity through

$$\hat{\alpha}_j^H = -2\frac{\bar{X}_j \tilde{\beta}_j^H}{s_j^2} \neq 0 \qquad (15)$$

if $\bar{X}_j \neq 0$. Notice that the probability that $\bar{X}_j = 0$ holds equals zero for observational studies. If $\bar{X}_j = 0$ holds for designed experiments, we can add $\delta \neq 0$ to $X_j$ initially to make $\hat{\alpha}_j^H \neq 0$.

If $X_j$ is selected, then the strong heredity is not violated. It is practically impossible that $\frac{\tilde{\alpha}_j^H}{s_j} - 2\frac{\bar{X}_j \tilde{\beta}_j^H}{s_j^2} = 0$. Even if it does happen, we can always introduce $\delta$ to adjust $\bar{X}_j$ to avoid $\hat{\alpha}_j^H = 0$.

Second, without loss of generality, if the quadratic term $X_j^2$ is not selected but some (or all) of the 2-factor interactions related to $X_j$ are selected, then $\tilde{\beta}_j^H = 0$ and some (or all) $\tilde{\gamma}_{jk}^H \neq 0$. Correspondingly,

$$\hat{\alpha}_j^H = \frac{\tilde{\alpha}_j^H}{s_j} - \sum_{\substack{k \neq j \\ 1 \leq k \leq p}} \frac{\bar{X}_k \tilde{\gamma}_{jk}^H}{s_j s_k} \qquad (16)$$

Following the same rationale, if $X_j$ is not selected ($\tilde{\alpha}_j^H = 0$ and the strong heredity is violated), then

$$\hat{\alpha}_j^H = - \sum_{\substack{k \neq j \\ 1 \leq k \leq p}} \frac{\bar{X}_k \tilde{\gamma}_{jk}^H}{s_j s_k} \neq 0 \tag{17}$$

if other $\bar{X}$'s are not equal to 0. If $X_j$ is selected, then the strong heredity is not violated.

Finally, when both the quadratic term $X_j^2$ and some (or all) 2-factor interactions are selected, we can follow the same proof and can conclude that the proposed hierarchical standardization guarantees the strong heredity. □

## Appendix 2: Algorithmic Procedure to Implement Hierarchical Standardization

Algorithmic procedure to implement hierarchical standardization

1: **Input:** Mean effects (X's) of the model and data
2: **Output:** A model of selected variables that satisfies the strong heredity
3: Step 1: Transform the main effects (e.g., using the mean-standard deviation transformation)
4: Step 2: Generate higher-order terms (2-factor interactions and quadratic terms) using the scaled main effects
5: Step 3: Implement a variable selection method (e.g., the lasso) and estimate the model using the generated main effects, 2fi's, and quadratic effects
6: Step 4: Back-transform coefficients to their original scales using Eq. (8)

## Appendix 3: An Illustrative Example of Hierarchical Standardization

We use an example in Setting 1 (see Table 1) to illustrate the hierarchical standardization. The variable selection problem that we consider is as follows.

$$Y = X_1 + X_2 + X_3 + X_1 X_2 + X_1 X_3 + X_2 X_3 + X_1^2 + X_2^2 + X_3^2 + \varepsilon \tag{18}$$

where $\varepsilon \sim N(0, 8^2)$. $X$ is a 10,400-by-65 matrix where 200, 200 and 10,000 are the numbers of rows in the training, validation and test datasets, respectively. The first 10 columns represent the main effects, the next 45 columns represent the 2fi's, and the final 10 columns represent the quadratic effects. Each of the first 10 columns is generated independently from the standard normal distribution. Each of the next 45 columns is the element-wise multiplication of the corresponding main-effect columns. Each of the final 10 columns is the square of the corresponding main-effect column. $Y$ is generated according to Eq. (18). In this sense, ideally we aim to identify Eq. (18) based on the $X$ and $Y$.

**Table 11** Lasso results for the illustrative example

| ME | $X_1$ | $X_2$ | $X_3$ | $X_4$ | $X_5$ | $X_6$ | $X_7$ | $X_8$ | $X_9$ | $X_{10}$ |
|---|---|---|---|---|---|---|---|---|---|---|
| | 0 | 0 | 1.0233 | −0.0608 | 0 | 0.1252 | 0 | 0 | 0 | 0 |
| 2FI | $X_1X_2$ | $X_1X_3$ | $X_1X_4$ | $X_1X_5$ | $X_1X_6$ | $X_1X_7$ | $X_1X_8$ | $X_1X_9$ | $X_1X_{10}$ | $X_2X_3$ |
| | 0.9058 | 0.0804 | 0 | 0 | 0 | 0 | 0 | 0 | 0 | 0.7116 |
| | $X_2X_4$ | $X_2X_5$ | $X_2X_6$ | $X_2X_7$ | $X_2X_8$ | $X_2X_9$ | $X_2X_{10}$ | $X_3X_4$ | $X_3X_5$ | $X_3X_6$ |
| | 0 | 0 | 0 | 0 | 0 | 0 | 0.3821 | 0.0419 | 0 | −0.2529 |
| | $X_3X_7$ | $X_3X_8$ | $X_3X_9$ | $X_3X_{10}$ | $X_4X_5$ | $X_4X_6$ | $X_4X_7$ | $X_4X_8$ | $X_4X_9$ | $X_4X_{10}$ |
| | 0.6411 | 0 | 1.3168 | 0 | 0.1662 | 0 | 0 | 0 | 0 | 0 |
| | $X_5X_6$ | $X_5X_7$ | $X_5X_8$ | $X_5X_9$ | $X_5X_{10}$ | $X_6X_7$ | $X_6X_8$ | $X_6X_9$ | $X_6X_{10}$ | $X_7X_8$ |
| | 0 | 0 | 0 | 0 | 0 | 0 | 0 | 0 | 0 | 0 |
| | $X_7X_9$ | $X_7X_{10}$ | $X_8X_9$ | $X_8X_{10}$ | $X_9X_{10}$ | | | | | |
| | 0 | 0 | 0 | 0 | 0 | | | | | |
| QE | $X_1^2$ | $X_2^2$ | $X_3^2$ | $X_4^2$ | $X_5^2$ | $X_6^2$ | $X_7^2$ | $X_8^2$ | $X_9^2$ | $X_{10}^2$ |
| | 0.3822 | 0 | 0 | 0 | 0.1222 | 0 | 0 | 0 | 0 | 0 |

**Table 12** Lasso results after hierarchical standardization for the illustrative example

| ME | $X_1$ | $X_2$ | $X_3$ | $X_4$ | $X_5$ | $X_6$ | $X_7$ | $X_8$ | $X_9$ | $X_{10}$ |
|---|---|---|---|---|---|---|---|---|---|---|
| | −0.0015 | −0.0284 | 1.2483 | −0.0518 | 0.0203 | 0.1157 | 0.0122 | 0 | 0.0281 | 0.0111 |
| 2FI | $X_1X_2$ | $X_1X_3$ | $X_1X_4$ | $X_1X_5$ | $X_1X_6$ | $X_1X_7$ | $X_1X_8$ | $X_1X_9$ | $X_1X_{10}$ | $X_2X_3$ |
| | 0.9651 | 0.0828 | 0 | 0 | 0 | 0 | 0 | 0 | 0 | 0.8249 |
| | $X_2X_4$ | $X_2X_5$ | $X_2X_6$ | $X_2X_7$ | $X_2X_8$ | $X_2X_9$ | $X_2X_{10}$ | $X_3X_4$ | $X_3X_5$ | $X_3X_6$ |
| | 0 | 0 | 0 | 0 | 0 | 0 | 0.427 | 0.044 | 0 | −0.258 |
| | $X_3X_7$ | $X_3X_8$ | $X_3X_9$ | $X_3X_{10}$ | $X_4X_5$ | $X_4X_6$ | $X_4X_7$ | $X_4X_8$ | $X_4X_9$ | $X_4X_{10}$ |
| | 0.6167 | 0 | 1.4174 | 0 | 0.1597 | 0 | 0 | 0 | 0 | 0 |
| | $X_5X_6$ | $X_5X_7$ | $X_5X_8$ | $X_5X_9$ | $X_5X_{10}$ | $X_6X_7$ | $X_6X_8$ | $X_6X_9$ | $X_6X_{10}$ | $X_7X_8$ |
| | 0 | 0 | 0 | 0 | 0 | 0 | 0 | 0 | 0 | 0 |
| | $X_7X_9$ | $X_7X_{10}$ | $X_8X_9$ | $X_8X_{10}$ | $X_9X_{10}$ | | | | | |
| | 0 | 0 | 0 | 0 | 0 | | | | | |
| QE | $X_1^2$ | $X_2^2$ | $X_3^2$ | $X_4^2$ | $X_5^2$ | $X_6^2$ | $X_7^2$ | $X_8^2$ | $X_9^2$ | $X_{10}^2$ |
| | 0.3614 | 0 | 0 | 0 | 0.1167 | 0 | 0 | 0 | 0 | 0 |

We apply the hierarchical standardization to the Lasso method. The training and validation sets have 200 observations each, and the test set has 10,000 observations. We supply the training $X$ and $Y$ to the R function glmnet. After using the validation $X$ and $Y$, the function returns the following coefficients (Table 11) that lead to the minimum mean squared error (MSE) on the validation set.

Table 11 clearly indicates a violation of the strong heredity. For instance, $X_1X_2$, $X_1X_3$, and $X_1^2$ are all active effects selected by the Lasso, but $X_1$ is not selected (coefficient is 0). Hence, we apply Eq. (8) to the coefficients in Table 11 and obtain the re-scaled coefficients as Table 12.

Table 12 shows the selected variables using the Lasso with hierarchical standardization. Noticeably, the strong heredity is not violated. We specifically investigate the coefficients of $X_1$, $X_1X_2$, $X_1X_3$, and $X_1^2$. After examination, $X_1$, $X_2$ and $X_3$ have sample means 0.03898826, $-$0.02594940 and $-$0.01980965, respectively. $X_1$, $X_2$ and $X_3$ have sample standard deviations 1.0283163, 0.9127104 and 0.9451362, respectively. The re-scaled coefficient of $X_1X_2$ (0.9651) comes from the original 2fi coefficient divided by the multiplication of the corresponding sample standard deviations, that is, $0.9058/(1.0283163 * 0.9127104) = 0.9651$. Similarly, the re-scaled coefficient of $X_1X_3$ (0.0828) comes from $0.0804/(1.0283163 * 0.9451362) = 0.0827$ (rounding issue), and the re-scaled coefficient of $X_1^2$ (0.3614) comes from $0.3822/1.0283163^2 = 0.3614$.

Finally, the re-scaled coefficient of $X_1$ ($-$0.0015) can be obtained through Eq. (8). That is,

$$0 - 2\frac{0.3822 \cdot 0.03898826}{1.0283163^2} - \left(\frac{-0.02594940 \cdot 0.9058}{1.0283163 \cdot 0.9127104} + \frac{-0.01980965 \cdot 0.0804}{1.0283163 \cdot 0.9451362}\right) = -0.0015 \quad (19)$$

As we can see, due to the back-transformation [Eq. (8)], main effects that are related to higher-order terms are forced to be active with non-zero coefficients in the model. Therefore, following the hierarchical standardization, the selected model can ensure the strong heredity.

## Results of Additional Robustness Check

This appendix presents results of additional robustness checks in Sect. 3.4.3. The extra simulation setups are shown in Table 13. Specifically, settings R4 through R6 correspond to the extension that the main effect coefficients are larger than those of 2fi's and quadratic terms. Settings R7 through R9 correspond to the extension of extra active main effects that do not appear in active higher-order terms. For instance, "3 + 1" under the setting R7 means we change one zero-coefficient main effect to non-zero (active). That main effect does not have associated active 2fi's or quadratic effects. In settings R4 through R9, we set coefficients that are similar to the setup in the main analysis.

Table 13 Setup for additional robustness checks

| Settings | R4 | R5 | R6 | R7 | R8 | R9 |
|---|---|---|---|---|---|---|
| # of main effects | 10 | 10 | 10 | 10 | 10 | 10 |
| # of non-zero main effects | 3 | 4 | 5 | 3 + 1 | 4 + 2 | 5 + 3 |
| # of non-zero 2fi's | 3 | 6 | 10 | 3 | 6 | 10 |
| # of non-zero quadratic effects | 3 | 4 | 5 | 3 | 4 | 5 |
| Coef of non-zero main effects | 3 | 5 | 5 | 1 | 1 | 5 |
| Coef of non-zero 2fi's | 1 | 3 | 3 | 1 | 3 | 3 |
| Coef of non-zero quadratic effects | 1 | 1 | 3 | 1 | 3 | 3 |
| $\sigma$ for the error term | 8 | 8 | 8 | 8 | 15 | 8 |
| Signal-to-noise ratio | 0.564 | 2.541 | 4.8 | 0.204 | 0.587 | 5.979 |

**Table 14** Results of additional robustness checks

|  | Settings | R4 | R5 | R6 | R7 | R8 | R9 |
| --- | --- | --- | --- | --- | --- | --- | --- |
| *Sensitivity* | | | | | | | |
| Our lasso | Mean | 0.804 | 0.969 | 1 | 0.786 | 0.955 | 1 |
|  | s.e. | (0.018) | (0.006) | (0.000) | (0.024) | (0.011) | (0.000) |
| Traditional lasso | Mean | 0.807 | 0.966 | 1 | 0.67 | 0.768 | 1 |
|  | s.e. | (0.018) | (0.006) | (0.000) | (0.029) | (0.019) | (0.000) |
| Our stepwise | Mean | 0.762 | 0.931 | 0.999 | 0.782 | 0.919 | 0.999 |
|  | s.e. | (0.017) | (0.009) | (0.001) | (0.017) | (0.010) | (0.001) |
| Traditional stepwise | Mean | 0.756 | 0.939 | 0.999 | 0.64 | 0.682 | 0.999 |
|  | s.e. | (0.018) | (0.008) | (0.001) | (0.020) | (0.014) | (0.001) |
| *Specificity* | | | | | | | |
| Our lasso | Mean | 0.684 | 0.568 | 0.461 | 0.732 | 0.639 | 0.464 |
|  | s.e. | (0.016) | (0.013) | (0.015) | (0.018) | (0.017) | (0.016) |
| Traditional lasso | Mean | 0.755 | 0.645 | 0.517 | 0.789 | 0.687 | 0.489 |
|  | s.e. | (0.016) | (0.015) | (0.017) | (0.019) | (0.017) | (0.018) |
| Our stepwise | Mean | 0.723 | 0.713 | 0.724 | 0.736 | 0.746 | 0.763 |
|  | s.e. | (0.010) | (0.009) | (0.010) | (0.010) | (0.009) | (0.010) |
| *Traditional stepwise* | Mean | 0.812 | 0.799 | 0.8 | 0.811 | 0.796 | 0.8 |
|  | s.e. | (0.010) | (0.009) | (0.010) | (0.010) | (0.010) | (0.010) |
| *MSE* | | | | | | | |
| Our lasso | Mean | 73.184 | 79.221 | 85.266 | 72.721 | 273.006 | 87.133 |
|  | s.e. | (0.436) | (0.625) | (0.82) | (0.347) | (1.937) | (0.854) |
| Traditional lasso | Mean | 73.126 | 79.12 | 85.073 | 72.702 | 272.094 | 87.142 |
|  | s.e. | (0.438) | (0.628) | (0.846) | (0.351) | (1.908) | (0.879) |
| Our stepwise | Mean | 86.95 | 89.098 | 89.665 | 87.579 | 318.854 | 90.847 |
|  | s.e. | (0.971) | (1.079) | (1.158) | (0.936) | (4.240) | (1.138) |
| Traditional stepwise | Mean | 86.778 | 88.821 | 89.623 | 87.883 | 318.778 | 90.604 |
|  | s.e. | (1.023) | (1.046) | (1.125) | (0.991) | (3.877) | (1.115) |

Our methods generate 100% maintenance of strong heredity for all settings R4 through R9. Table 14 gives the results of additional robustness check. For each setting, comparing our method to the traditional method, we observe comparable sensitivity, specificity and MSE. In this sense, the hierarchical standardization can guarantee the strong heredity without sacrificing estimation or prediction performance too much (Fig. 3).

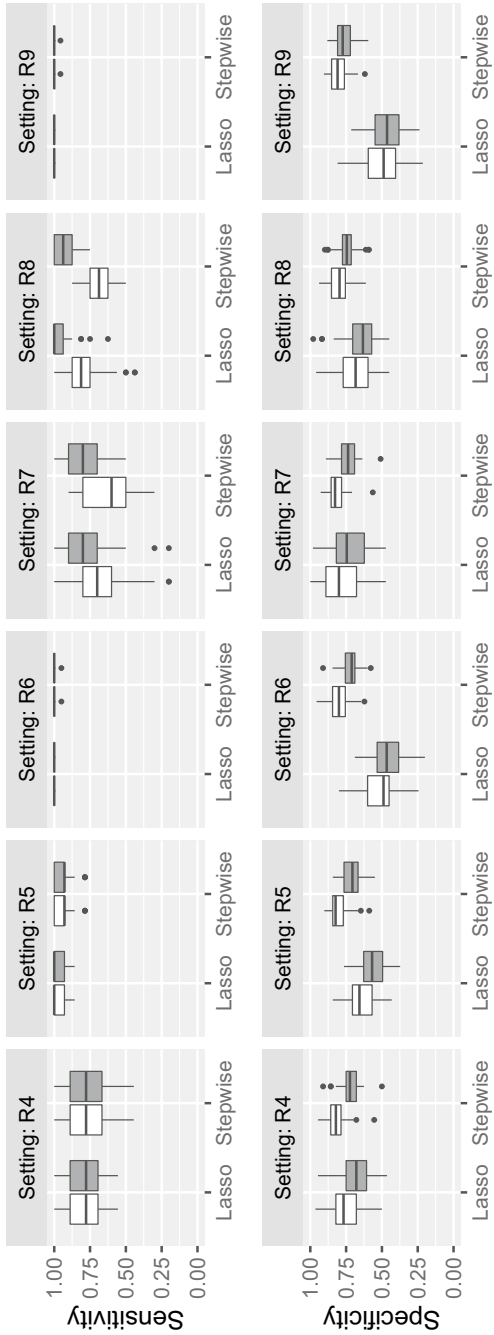

**Fig. 3** Boxplots of the sensitivity and specificity of our (grey boxes) and traditional (white boxes) methods, for additional robustness check